\documentclass[reprint,amsmath,amssymb,apr,aip,onecolumn]{revtex4-2}
\usepackage{graphicx}
\usepackage{graphics}
\usepackage{algorithm,algorithmic}
\usepackage{mathptmx}
\usepackage{times}
\usepackage{amsmath}
\usepackage{amssymb}
\usepackage{dcolumn}
\usepackage{color}
\usepackage{bm}
\usepackage{units}
\usepackage{booktabs}
\usepackage{multirow}
\usepackage[version=4]{mhchem}
\usepackage{siunitx}[=v2]
\usepackage{charter}
\usepackage[acronym]{glossaries}

\newacronym{pcm}{PCM}{phase-change memory}
\newacronym{imc}{IMC}{in-memory computing}
\newacronym{mvm}{MVM}{matrix-vector multiplication}
\newacronym{sem}{SEM}{scanning electron microscope}
\newacronym{tem}{TEM}{trasmission electron microscope}
\newacronym{mac}{MAC}{multiply-accumulate}
\newacronym{snr}{SNR}{signal-to-noise ratio}

\begin{document}
\title{Analogue Phase Change Computational Memory with High Precision Reads and Energy Efficient Writes}

\author{Ghazi Sarwat Syed} \affiliation{IBM Research -- Europe, S\"{a}umerstrasse 4, 8803 R\"{u}schlikon, Switzerland}
\author{Loris Coccia}\affiliation{IBM Research -- Europe, S\"{a}umerstrasse 4, 8803 R\"{u}schlikon, Switzerland}
\author{Vara Prasad Jonnalagadda}\affiliation{IBM Research -- Europe, S\"{a}umerstrasse 4, 8803 R\"{u}schlikon, Switzerland}
\author{Antonio Massimiliano Mio}\affiliation{Consiglio Nazionale delle Ricerche, 95121 Catania, Italy}
\author{Asit Ray} \affiliation{IBM Research--  Yorktown Heights,  10598 NY, USA}
\author{Matthew BrightSky} \affiliation{IBM Research--  Yorktown Heights,  10598 NY, USA}
\author {Abu Sebastian} \affiliation{IBM Research -- Europe, S\"{a}umerstrasse 4, 8803 R\"{u}schlikon, Switzerland} 

\maketitle
\noindent\textbf{Resistive memory technologies offer a compelling advantage for in-memory computing. However, realizing a device architecture that simultaneously achieves high computational precision, efficiency, and density has remained elusive due to inherent trade-offs among these performance metrics. Here, we introduce a compact phase-change memory device architecture that combines an ultra-confined active switching volume for enhanced electro-thermal efficiency with a non-insulating thin film that suppresses temporal conductance fluctuations. We analytically model and back-end integrate these devices into crossbar arrays. Even with conventional, undoped phase-change materials, the device architecture enables a computational precision approaching 6 bits, low conductance values below $50~\mu\text{S}$ with an adequate conductance window, and a viable pathway toward sub-$100~\mu\text{A}$ programming currents under nominal operating voltages.}\\

\noindent{Keywords: Phase Change Materials, Memory Devices, In-Memory Computing}

%--------------------------
% INTRODUCTION
%--------------------------

\section*{INTRODUCTION}

% What are the challenges
\noindent \Gls{pcm} has traditionally been developed for storage class memory and embedded memory applications\cite{Y2020kimTED,Y2020cappellettiJPD}. Recent research efforts, however, have shown significant potential of this technology for analogue \gls{imc} particularly for performing deep neural network inference\cite{Y2022khwaISSCC, Y2023legalloNatElec,Y2022LanzaScience}. In deep neural networks, \gls{mvm} operations dominate computation and can be implemented using crossbar arrays of synaptic unit cells, each containing one or more \gls{pcm} devices alongside selector devices such as field-effect transistors\cite{Y2025Syed2ChemicalReviews,Y2020sebastianNatNano}. 
Analogue \gls{imc} with \gls{pcm} offers the promise of unprecedented compute density and weight capacity compared to approaches based on static random access memory. Realizing this potential, however, requires improvements in the integration density of \gls{pcm} devices, which hinges on reducing the write or programming current that determines the size of selector devices and, consequently, unit cells. Another essential device-level feature is maintaining low conductance across all \gls{pcm} phase configurations to reduce energy consumption during \gls{mvm} operations and to mitigate non-idealities, such as parasitic voltage drops due to crossbar wire resistance. Additionally, intrinsic temporal non-idealities, specifically conductance drift and read noise, limit the achievable numerical precision during programming and computation. Suppressing these non-idealities is therefore critical for high-precision \gls{imc}.\\

% How it being solved
In efforts to enhance \gls{pcm} devices for analogue \gls{imc}, much research has concentrated on optimizing the composition of phase-change materials\cite{Y2019zhangNatRevMat,Y2017SyedMST,Y2021DragoniNanoscale}. More recently, significant progress has been made in exploring phase-change superlattices\cite{Y2025PriliAMI,Y2019dingScience,Y2021khanScience}. Although such approaches have demonstrated promise in micron-scale devices, the most effective and industrially proven strategy—one that accounts for process integration compatibility—remains the reduction of the material volume that undergoes the phase transition, thereby lowering the required write current\cite{Y2014BoniardiIEDM}. However, analogue \gls{imc} requires a continuum of conductance states at low absolute conductance values. Simply reducing the volume of phase-change material by reducing the span of phase change channel to achieve can degrade analogue behavior, as well as increase fabrication complexity, and—at extreme scales—promote elemental segregation under strong geometrical confinement. On the other hand, to suppress the non-idealities, a promising alternative is to project the amorphous volume of the phase-change material onto a non-insulating projection layer (liner), which effectively dictates the read-out conductance\cite{Y2015koelmansNatComm,Y2023SyedIEDM}. Projection efficiency is to be highest when the liner is fully and uniformly aligned in parallel with the active phase-change material. This configuration is difficult to achieve in conventional device architectures, such as mushroom-type vertical devices\cite{Y2021sarwatAFM}, which are primarily optimized for high integration density rather than optimal projection alignment.  \\

% What have we achieved 
In this article, we introduce and demonstrate a phase-change computational memory concept that elegantly fulfills both requirements. While retaining a compact vertical geometry, our design minimizes the active phase-change volume by reducing both the phase-change material thickness and its contact area with the bottom electrode\cite{Y2025SyedIEDM}, while simultaneously enhancing projection through a continuous physical connection to a non-insulating projection liner. We characterize these devices on a back-end-integrated crossbar array with MOSFET selectors. The devices enable low programming currents and conductance values, suppresses temporal conductance fluctuations, and remains compatible with conventional fabrication processes, making it ideally suited for analogue IMC applications.

%--------------------------
% THE CONCEPT
%--------------------------
\section*{Device concept}

\begin{figure}[h!]
    \includegraphics[width=1\textwidth]{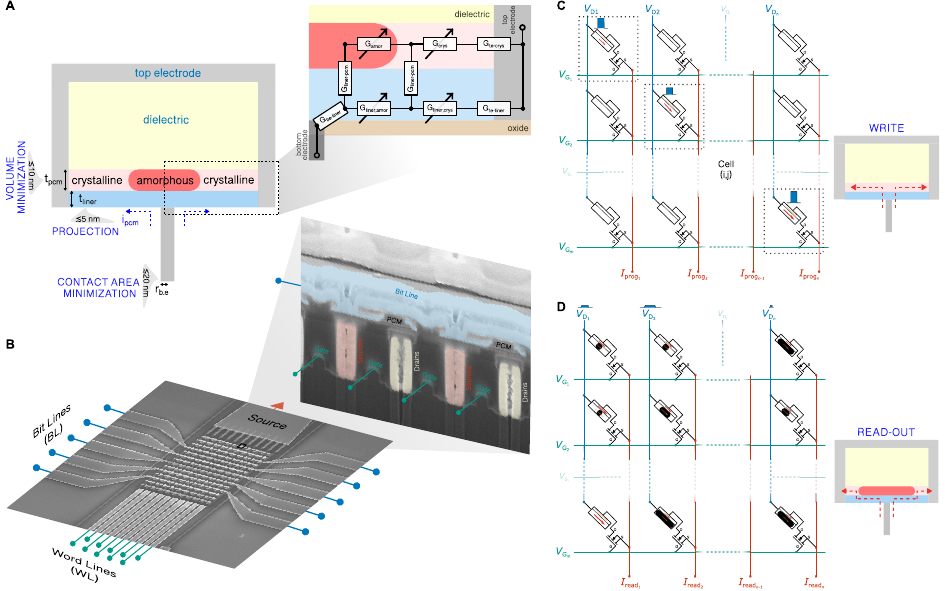}
    \caption{\textbf{Device concept}. (A) Cross-sectional schematic of the proposed computational phase-change memory device. The device uses an ultra-thin ($\lesssim \unit[10]{\rm nm}$) phase-change layer and a projection liner ($\lesssim \unit[5]{\rm nm}$) positioned between the dielectric and the bottom electrode. During programming or reading, current flows radially from the bottom electrode across either the phase-change layer or the liner to the edge-contacted top electrode. The device achieves high compute -efficiency (via reduced volume and contact area), -precision (through liner suppressing temporal fluctuations), and -density (from vertical geometry and efficient programming).  The inset shows the equivalent electrical circuit model, represented by lumped elements, including the contact interfaces. Here, $G_\mathrm{ele}$ represents the conductance contributions from the various interfacial and bulk elements which are critical for understanding and designing the device behavior. (B) \Gls{sem} micrograph of an crossbar array; each cross-point contains an nMOS selector in series with a phase-change device. Inset: \gls{tem} micrograph of a single device.  (C) Illustration of programming (write) operation: devices are programmed to analogue states (phase configurations), with current flowing through the higher-conductance phase-change segment, generating Joule heating and triggering the amorphization or crystallization.  (D) Illustration of readout operation: devices of varying conductance (phase configurations) can be read individually or used collectively for \gls{mac} operations. In this mode, current bypasses the amorphous region and flows through the non-drifting projection liner segment parallel to the liner.}
    \label{fig:1}
\end{figure}

% Explain the device strucutre, also discussing footprint
\noindent The proposed device architecture, illustrated in Figure \ref{fig:1}A, consists of an ultra-thin phase-change material film sandwiched between a confined bottom electrode and a dielectric spacer.  Between the bottom electrode and the phase-change layer lies a ultra-thin projection liner, which spans the full lateral extent of the device, corresponding to a radius of \( r_\text{pcm} \). Electrical contact is made laterally via a top electrode that wraps around the edge of the phase-change film, enabling current to flow outward from the central bottom electrode. A short, high-current RESET pulse amorphizes a region of radius \( u_{\text{a}} \) of the crystalline phase-change material adjacent to the bottom electrode. The overall device conductance is determined by the geometry and the extent of the amorphous region given by its radius \( u_{\text{a}} \) (colored red), which is electrically connected in parallel with the projection liner. Geometrically, both the amorphous region and the liner take the form of cylindrical discs, benefiting from the physics of ultra‑confinement. Furthermore, this configuration yields an inverse logarithmic dependence of conductance on \( G_\text{device} \propto 1/\log(u_{\text{a}}) \). Consequently, in this device concept the conductance is highly sensitive to small variations in the amorphous radius when $u_{\text{a}}$ is close to $r_{\text{BE}}$, whereas this sensitivity decreases for larger amorphous radii. Moreover, the lowest achievable conductance state scales with the ratio \( r_{\text{pcm}}/r_{\text{be}} \). Consequently, reducing the bottom-electrode radius permits a proportional reduction in device foot-print, which aligns naturally with advances in the lithographic scaling of minimum feature sizes. \\

% Explain the projection science
The device in a RESET state can be modeled (see inset of Figure \ref{fig:1}A) with the following conductive components: $G_{\text{amor}}$ (conductance arising from the amorphous volume), $G_{\text{crys}}$ (conductance from the crystalline volume), $G_{\text{liner,amor}}$ (conductance from the liner underneath the amorphous volume) and $G_{\text{liner,crys}}$ (conductance from the liner underneath the crystalline volume). Two additional conductance components determine the device functionality: $G_{\text{liner-pcm}}$ (contact resistance between the liner and the phase-change material) and $G_{\text{be-liner}}$ (contact resistance between the liner and the bottom electrode). The electrical conductivities of the phase-change and liner materials satisfy $\sigma_{\mathrm{amor}} < \sigma_{\mathrm{liner}} < \sigma_{\mathrm{crys}}$, where the conductivity contrast is defined via $\sigma_{\mathrm{liner}} = \alpha\,\sigma_{\mathrm{crys}}$ for a tunable scaling parameter $\alpha \in (0, 1)$. Together with the geometrical parameters (film thickness), this conductivity contrast enables optimization of the memory window and projection efficacy. In the limit of a small $G_{\text{be-liner}}$, and for arbitrary values of $G_{\text{liner-pcm}}$, the minimal achievable conductance approaches $G_{\text{be-liner}}$. We are able analytically describe these device characteristics by incorporating the relevant bulk and interfacial material properties, as well as geometric parameters. The model accurately captures both experimental observations and finite-element method (FEM) simulations. We find that the contact resistance between the liner and the phase-change material plays a dominant role in determining projection efficacy\cite{kersting2020state}. An increase in this resistance degrades analogue programmability and limits the correction of the amorphous-phase non-idealities, as will be discussed in a following section.

%------------------------------------------------------------------------------------------------
% EXPERIMENTAL VALIDATION
%------------------------------------------------------------------------------------------------
\section*{Experimental validation}

% How does a crossbar array function 
\noindent Prototype devices were fabricated using the conventional, undoped phase-change material \ce{Ge2Sb2Te5} (GST) with a thickness approaching \unit[5]{nm} and an amorphous-carbon (aC) based liner of \unit[3]{nm}. The devices were integrated with n-MOS transistors serving as selectors, with bottom electrodes formed from chemical mechanical polished metal nitride (\ce{Ti_xN_y}) of $r_\text{be}$ \unit[20]{nm}. Additionally, a $10 \times 8$ array of these devices was designed and fabricated, incorporating designated word lines for transistor control, along with separate drain and source lines for programming and read-out operations (see Figure \ref{exfig:1}B-D). To further minimize the bottom-electrode to \gls{pcm} contact area, devices with unplanarized bottom electrodes penetrating directly into the device stack were also fabricated.\\

\begin{figure}[h!]
\includegraphics[width=\textwidth]{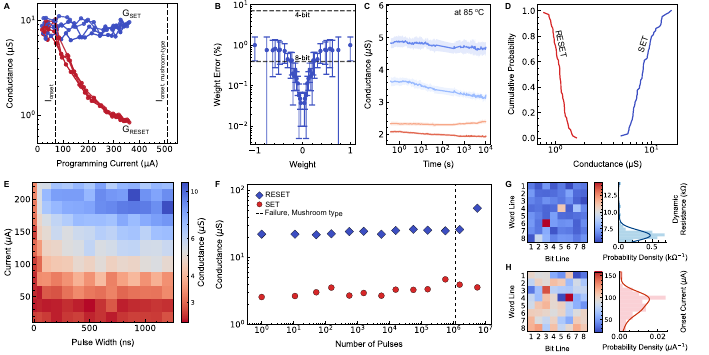}
\caption{\textbf{Programming characteristics}. (A) A plot illustrating the programming curves of a device showing a notable reduction in programming currents. (B) A plot showing the weight error obtained from mapping synaptic weight values onto the analogue conductance states of the device. The error bars in represent one standard deviation. Also indicated are the 4-bit and 8-bit effective quantization noise. (C) The temporal evolution of conductance for a two representative states of the device at elevated temperature (\unit[85]{$^\circ$C}), showing stable retention behavior. (D) Reliability plot showing the complementary cumulative distribution of RESET conductance values and the cumulative distribution of SET conductance values measured across 64 devices in the crossbar array, obtained from single-pulse programming. (E) Measurement of the crystallization speed in the device using box-type SET pulses of varying pulse amplitude and duration. The color map shows the device conductance as a function of the pulse parameters. (F) A typical endurance plot, where a device is repeatedly stressed by bursts of RESET pulses. It can be seen that it is possible to achieve reliable reversible switching between SET and RESET states in excess of 100 million cycles. (G) Heat map and corresponding distribution illustrating the dynamic ON-resistance values of 64 devices. Higher $R_{\rm ON}$ enhances Joule heating efficiency, resulting in a reduced onset current. (H) Heat map and corresponding distribution of the onset current across the 64 devices.}
\label{fig:2}
\end{figure}

% Programming current
We programmed the devices using successive RESET pulses with increasing amplitude, which melt-quench and amorphize the phase-change material, followed by SET pulses that crystallize the amorphous regions created by any of the RESET pulses. The device conductance was measured at a \unit[0.1]{V} bias after each SET or RESET pulse, generating a programming curve based on the amplitude of the current programming pulse needed. Figure \ref{fig:2}A shows four representative programming curves from the device, together with reference lines indicating the SET conductance and onset current of a conventional mushroom device with \unit[80]{nm} GST. The device achieves a continuum of analogue states, with a reduction in the onset current, from approximately \unit[500]{$\mu$A} in the mushroom-type device to \unit[90]{$\mu$A}. Note that we define the onset current as the point at which the device conductance begins to decrease. This point corresponds to the onset of melting and helps avoid ambiguity in resistance scaling behavior, which can depend on the device geometry. In addition to these improvements in programming, the devices also achieve over 10x lower SET conductance , while preserving a large programming window of more than 10x. These results validate that physical scaling of the device dimensions and contacts leads to strong electro-thermal confinement, resulting in more efficient Joule heating within the phase-change film and a reduction in conductance values.\\

% discuss characteristic of each state: weight error, retention, crystallization speed, endurance
We are able to accurately and reliably program the devices to target conductance states using a standard iterative programming scheme. Figure \ref{fig:2}B illustrates the programming error associated with encoding synaptic weights. In this experiment, eighteen target conductance states (\(G_\text{Target}\)) are programmed ten times within a $\pm 2\%$ margin. The weight error is defined as
\(|G_\text{Target} - G|/G_\text{max}\),
where \(G\) is the device conductance measured after convergence, and \(G_\text{max}\) is the maximum reliably programmable unit-cell conductance. Our ability to program within a narrow margin is enabled by the suppressed conductance fluctuations (read noise), which we will discuss shortly. Furthermore, we find that the programmed states exhibit excellent retention characteristics. The programmed conductance traces exhibit stable retention behavior across the measured conductance range, even at elevated temperatures (e.g., 85 $^\circ$C, as shown in Figure \ref{fig:2}C).  Furthermore, we find that even with single shot (open-loop) programming, the RESET and SET conductance states are well separated across the crossbar array, as shown by the complementary cumulative RESET distribution and the cumulative SET distribution (see Figure~\ref{fig:2}D). Additionally, the RESET states show a characteristic crystallization behavior that depends on the amplitude and duration of the SET pulse. To demonstrate this, we applied box-type SET pulses with fixed falling and trailing edges of \unit[8]{ns}, but varying amplitude and width (see Figure \ref{fig:2}E). The color map shows the device conductance as a function of the pulse parameters, revealing a SET speed of approximately \unit[900]{ns}. Figure~\ref{fig:2}F demonstrates reliable programming of the device under a burst of RESET pulses over 1 million cycles. This endurance exceeds the requirements for \gls{imc} applications, where frequent reprogramming is unnecessary. Notably, it represents a $100\times$ improvement over undoped-GST compositions, which we attribute to the physical confinement of the phase-change material. This confinement likely suppresses elemental segregation mechanisms.\\

% discuss array level measurements on programming
Figure~\ref{fig:2}G–H presents the characterization measurements across multiple devices in the crossbar array. 
The left panel of Figure~\ref{fig:2}G shows a heat map of the dynamic on-resistance ($R_{\text{ON}}$) measured across 64 devices. A higher $R_{\text{ON}}$ corresponds to high Joule heating efficiency. Notably, we find that compared to the mushroom-type device, $R_{\text{ON}}$ is improved by $\sim$10x. The right panel shows the corresponding statistical distribution in $R_{\text{ON}}$, highlighting the device-to-device variability. Similarly, the left panel of Figure~\ref{fig:2}H illustrates the heat map of the programming onset current ($I_{\text{Prog}}$). Notably, the higher $R_\text{ON}$ values leads to lower $I_\text{Prog}$ values in the devices, in part explaining the improved programming efficiencies. The right panel presents $I_\text{Prog}$ distribution across the same 64 devices. Note that both parameters exhibit a tight distribution. State-dependent threshold voltages for the SET transition were similarly characterized across the array, featuring forming-free operation at sub-\unit{2}{V} voltages. \\

\begin{figure}[h!]
\includegraphics[width=\textwidth]{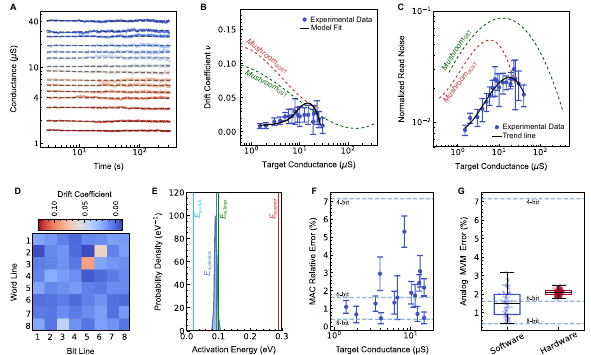}
\caption{\textbf{Read-out characteristics}. (A) Conductance versus time traces of many programmable states in a device. The data is fitted with the conductance drift equation, $G(t) = G(t_0) (t/t_0)^{-\nu_\text{eff}}$ (black dotted traces). (B) A plot comparing the proposed computational memory device with undoped and doped mushroom-type devices on state-dependent conductance drift coefficients. The black trace is the fit to the data using the analytical model. (C) Comparison of the state-dependent normalized read noise of the proposed computational memory device with that of undoped and doped mushroom-type devices. The normalized read noise is calculated as the standard deviation of the residual raw conductance traces after subtraction of the drift component, divided by the target conductance. Data in (B) and (C) are averaged over 10 repetitions per iteratively programmed state; error bars indicate one standard deviation. (D) Heat map illustrating the spatial distribution of drift coefficients for full RESET states across 64 devices in the crossbar array. (E) Distribution of activation energies extracted from measurements of 27 devices in the RESET state. (F) State-dependent relative error of individual multiply–accumulate (MAC) outputs, where each MAC corresponds to the scalar dot product between one crossbar row and the applied input vector, with all devices participating in the MAC programmed to the same target conductance state. Measurements are shown for different target conductance states, together with the error limits associated with different digitally quantized precisions. (G) Analogue matrix–vector multiplication errors, quantified as the normalized ($l_2$) difference between the complete measured and ideal output vectors, comparing accumulation performed in software with full hardware accumulation on the crossbar array.
}    
\label{fig:3}
\end{figure}

% Temporal behavior
We evaluated additional critical parameters for \gls{imc}, namely the temporal conductance variations caused by conductance drift and read noise. Drift refers to the progressive decrease in conductance within the RESET state, while read noise constitutes the random, typically $1/f$ fluctuations superimposed on the resting conductance value. Together, drift and read noise determine the weight error and, consequently, the accuracy of \gls{mvm} operations. Figure~\ref{fig:3}A shows the conductance versus time for multiple non-volatile states in a device. By fitting these data with the standard drift model, we extract the effective state-dependent drift coefficients ($\nu_\text{eff}$). Evidently, conductance drift is significantly suppressed in these devices. For RESET states, the drift coefficients are reduced by up to a factor of 20 relative to mushroom-type GST devices, highlighting the effectiveness of the projected-cell design. Similarly, for intermediate states, the devices exhibit drift coefficients lower than those of doped GST. The latter is a current state-of-the-art material engineered to suppress conductance values to the scale we otherwise achieve through our device geometry. Figure \ref{fig:3}B shows a comparison of the state-dependent drift coefficients across different conductance states. For this analysis, each target conductance state ($G_{\text{T}}$) was programmed using the iterative programming scheme, and its conductance versus time behavior was fitted over ten separate trials.  We can accurately model the state-dependent conductance of our devices using the analytical description described earlier. This state dependency originates from the underlying phase configurations. 
For larger values of \(u_\text{a}\), the liner begins to dominate the resistor network, leading to decreasing $\nu_\text{eff}$. As with improvements in the drift behavior, we find that our good projection efficacy significantly suppresses temporal fluctuations, leading to a reduction in read noise ($\sigma_{\text{G}}$), which was extracted by subtracting the model-fitted drifting conductance from each individual measurement and computing the standard deviation of the residual noise. Figure~\ref{fig:3}C plots the state-dependent \gls{snr}, defined as $\sigma_{\text{G}}/G_{\text{T}}$, for various target conductance states that we have iteratively programmed. Notably, the SNR improves by a factor of $4\times$ compared to to mushroom-type GST devices. \\

% discuss array level measurements on read-out

In Figure~\ref{fig:3}D, we plot the heat map of the drift coefficients in the full RESET states of 64 devices in the crossbar array. Notably, the drift coefficient are suppressed in all devices. Conductance change due to ambient temperature fluctuations is another important device state-variable that impacts temporal stability of the states. This effect is more prevalent in the RESET states, due to the non-trivial activation energies ($E_\mathrm{a}$) for charge transport in the amorphous volume. For \unit[$\sim$5]{nm }GST we measure $E_\mathrm{a,amor} = \unit[0.29]{eV}$. Projection is expected to suppress the temperature dependence of the device, with the effective activation energy of the device, $E_\mathrm{a,eff}$, approaching that of the liner (i.e., $E_\text{a,liner}, \unit[0.10]{eV}$). To verify this, we measured $E_\mathrm{a,eff}$ across 27 randomly selected devices in the crossbar array. For each device, we created a RESET state and measured its conductance at different ambient temperatures. The resulting data were fitted to the Arrhenius equation, that scales conductance exponential to $E_\mathrm{a}$. Figure~\ref{fig:3}E illustrates the distribution of $E_\mathrm{a,eff}$, evidently showing a mean value close to $E_\text{a,liner}$.\\

% discuss compute precison 
Collectively, these results indicate that these devices can improve computational precision in analogue arithmetic operations. To validate this, we performed fully hardware-implemented \gls{mvm} experiments. We first investigated the state-dependent contributions to computational accuracy. To this end, multiply--accumulate (MAC) operations were performed by programming all devices connected to a common bit-line to conductance states sampled from a uniformly distributed weight matrix $W$. A close loop iterative programming procedure with $1\%$ error tolerance was used to map the weights in conductance states. A total of 250 independent measurements, each consisting of a constant voltage amplitude of \unit[0.1]{V} vector, were applied to the word lines. The resulting device currents, representing the scalar multiplication operations, were accumulated along the shared bit-line. Note that this measurement approximates the pulse-width modulation scheme commonly used in AIMC, where input values are encoded as constant-amplitude voltage pulses. The MAC error was defined as $\frac{y_{\mathrm{ideal}}-y_{\mathrm{experimental}}}{y_{\mathrm{ideal}}}$, which quantifies the linear error arising from the hardware implementation, including both the PCM devices and the associated interconnect network. The state-dependent MAC errors obtained from 250 independent measurements are shown in Figure~\ref{fig:3}F. Notably, conductance states exhibiting strong projection achieve computational precision exceeding 6 bits. In contrast, intermediate conductance states closer to the SET state exhibit larger errors. This behavior is consistent with the analytical model and is expected to be further mitigated through electrode engineering. Next, we evaluated complete \gls{mvm} operations by programming the crossbar array with a uniformly sampled weight matrix $W$. The resulting computational precision is shown in Figure~\ref{fig:3}G. Notably, the hardware achieves more than 5 bits of effective computational precision. These results are further compared with an emulator incorporating the experimentally observed device non-idealities. Importantly, this level of precision represents a non-trivial improvement over state-of-the-art computational memory systems, which typically report 3--4 bits of effective computational accuracy\cite{Y2023SyedIEDM}.

%------------------------------------------------------------------------------------------------
% DISCUSSION
%------------------------------------------------------------------------------------------------
\section*{Discussion and Outlook}

\begin{figure}[h!]
\includegraphics[width=0.60\textwidth]{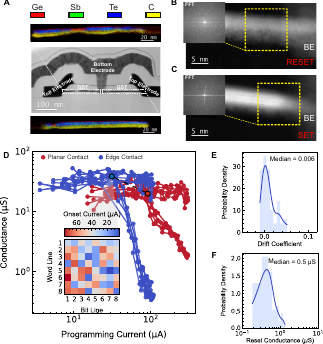}
\caption{\textbf{Prospects for further contact area minimization}. (A) Bright-Field scanning transmission electron microscopy (BF-STEM) micrograph of a device with a penetrating bottom electrode, together with the corresponding elemental composition maps.
(B) HAADF-STEM image of a device in the RESET state, showing an amorphous region adjacent to the bottom electrode (BE). The corresponding FFT pattern confirms the amorphous character of the region. (C) HAADF-STEM image of a device in the SET state, showing a crystalline region adjacent to the BE. The corresponding FFT pattern confirms the crystallinity in the region. (D) Comparison of programming curves between planar- and edge-contacted devices, showing improved performance when the bottom electrode fully penetrates the device stack. Inset: heat map of the onset programming currents measured across 64 devices in the array. (E) Distribution of drift coefficients measured across 64 devices, highlighting significant drift reduction. (F) Corresponding RESET conductance values across the measured devices.}
\label{fig:4}
\end{figure}

\noindent The geometry of our proposed device introduces an additional design parameter for the placement of the bottom electrode to minimize contact area. When the bottom electrode fully penetrates the phase-change material–liner stack, the contact geometry transitions from a planar circular interface to a cylindrical sidewall interface. Based on our devices parameters, this configuration increases the local power density at the contact, enhancing Joule heating efficiency, and raises the effective thermal resistance for heat dissipation, improving thermal confinement; thereby enhancing the programming efficiency. For the geometric dimensions of our devices, where the heater radius is comparable to the phase-change material thickness, this reduction in contact area approaches $1/2$. To validate this design strategy, with the same material stack, we fabricated and integrated devices in which the bottom electrode penetrates completely through the stack, establishing an edge-contact configuration similar to the contact made by the side electrode (see Figure~\ref{fig:4}A (i)).  Figures~\ref{fig:4}B-C show HAADF-STEM micrographs of a cycled device in a RESET state, and in SET state, respectively. Crucially, we observe clear evidence of amorphized and recrystallized volumes in the devices, which is consistent with the melt-quench and crystallization behavior of GST, as also confirmed by the corresponding Fast-Fourier Transform (FFT) patterns. Also note that the structural integrity of the confined bottom electrode and the stoichiometry of the phase-change material remain intact.  Atomic-resolution HAADF-STEM imaging of the phase change material channel in both the RESET and SET states, reveals a highly ordered layered structure. Distinct pseudo van der Waals (vdW) gaps, appearing as dark horizontal lines parallel to the phase change channel, indicate a pronounced texture with preferential alignment of the crystalline grains. The measured distances between neighboring vdW gaps are approximately 1.05~nm and 1.35~nm, corresponding to the characteristic thicknesses of Sb$_2$Te$_3$ and Ge$_1$Sb$_2$Te$_4$ structural blocks, respectively. The observation of the same layered motifs in both conductance states suggests that the overall crystalline framework remains preserved during electrical switching. The structural models superimposed on the HAADF-STEM images corroborate the identification of these layered building blocks and their ordered stacking within the phase-change material channel. \\

Moreover, as shown in Figure~\ref{fig:4}D, the programming efficiency itself is significantly improved. The programming current drops by nearly half to \unit[50]{$\mu$A}, while low conductance states are reliably maintained across all programmable levels. For the 64 devices in the crossbar array, the inset presents a heat map highlighting the spatial distribution of the programming onset current. Figure~\ref{fig:4}E–F show the corresponding distributions of the drift coefficients and RESET conductance values. 
The extracted drift coefficients exhibit a median value of $\nu_{\text{eff}} = 0.006$, indicating limited temporal drift across devices. These results demonstrate that the characteristic signatures of good projection efficacy are consistently preserved in these device structures. Similarly, further decreasing the thickness of the GST film can also provide a means to more significantly decrease the device conductance\cite{Y2025SyedIEDM}.  Together, these avenues validate improvements that can be enabled to our proposed device concept.\\

We would also like to point out a specific feature of the projection is that it reduces the apparent memory window, typically defined as the resistance ratio $R_{\text{RESET}}/R_{\text{SET}}$. In \gls{imc}, however, the quantities of primary interest are conductances rather than resistances. The resistance-to-conductance mapping acts like a squashing function, already compressing large resistance values. Moreover, the number of reliably usable discrete levels scales with the conductance separation (not their ratio) and inversely with the degree of temporal fluctuations, i.e., it is proportional to $\frac{G_{\max}-G_{\min}}{\sigma_{\text{total}}}$, where $\sigma_{\text{total}}$ represents the temporal conductance variability arising from both read-noise and conductance drift\cite{sarwat2023mechanism}. Consequently, variations in $G_{\min}$ alone have a less significant impact on the effective operational window for mapping synaptic weights. As an example, conventional mushroom-type \gls{pcm} devices may exhibit memory windows exceeding $10^3$, yet still accommodate fewer stable discrete levels than our proposed devices. This improvement arises primarily from the suppressed temporal variations achieved in our architecture\cite{Y2023LiAEM}. As next steps, exploring alternative material compositions presents promising opportunities for performance enhancement, such as ultra-thin superlattice and heterostructure films that offer natural confinement properties, and GeSb, nanocomposite films for lower threshold voltages and improved retention\cite{Y2021kwonNanoletters,Y2006chenIEDM}. For commercial integration, achieving wafer-scale film uniformity is essential, especially because the material properties are highly sensitive to thickness variations in sub-\unit[10]{nm} films. In this proof-of-concept study, we already demonstrate the fabrication feasibility using the industry-preferred sputter deposition method. In cases where tighter control over film thickness is required, especially for thicknesses approaching sub-\unit[3]{nm}, more conformal deposition techniques, such as atomic and chemical layer deposition may become required. These deposition schemes are already widely optimized in the fabrication of confined-type devices. We would also like to highlight that while some advanced materials and device structures show indications of similar improvements, such structures largely limit analogue capability due to the small material volume required for melting and/or pose challenges for large-scale integration with back-end-of-line processes.

%------------------------------------------------------------------------------------------------
% CONCLUSIONS
%------------------------------------------------------------------------------------------------
\section*{Conclusions}
\noindent In summary, we have introduced a \gls{pcm} device concept for analogue in-memory computing that achieves low programming current, low conductance values, and high compute precision, while being well-suited for large-scale back-end-of-line integration. We have experimentally demonstrated this concept by fabricating prototype CMOS integrated crossbar array with the standard GST material, which can reversibly switch at programming currents around \unit[$<$100]{$\mu$A} while maintaining a conductance values below \unit[50]{$\mu$S}, and suppressed drift and read noise, allowing more than 5-bit compute precision. Further enhancements may be possible by reducing the phase-change material thickness, minimizing the electrode contact area, and using optimized phase-change materials. These results mark a significant advancement toward enabling low-power, high-density resistive memory based in-memory computing.

%------------------------------------------------------------------------------------------------
% MATERIALS AND METHODS
%------------------------------------------------------------------------------------------------
\section*{MATERIALS AND METHODS}

\small

\subsection*{Device fabrication}
\noindent Deposition of the PCM stack on \unit[180]{nm} node n-MOS front-end-of-line wafers was initiated by in situ etching (ISE) of the heater electrode, drain-contact region, source, gate, and ground pads using Ar plasma for \SI{1}{min} at \SI{100}{W}. A \SI{4}{nm} doped amorphous carbon (a-C) projection layer containing $<\SI{20}{sccm}$ Ar dopant with H$_2$ was subsequently deposited, followed by the \emph{in situ} deposition of a \SI{5}{nm} Ge$_2$Sb$_2$Te$_5$ (GST225) layer by \SI{50}{W} DC magnetron sputtering. Finally, a \SI{6}{nm} SiO$_2$ capping layer was deposited by \SI{600}{W} RF sputtering. The cell structures were defined in a e-beam lithography step. A \SI{70}{nm}-thick negative-tone hydrogen silsesquioxane (HSQ) resist layer was spin-coated and exposed, and the patterns were transferred into the underlying layers by ion milling. Immediately after etching, and to minimize oxidation of the PCM sidewalls, the wafer was transferred to a sputtering system, where the exposed sidewalls were cleaned using Ar plasma and subsequently encapsulated with a \SI{50}{nm} W layer. For the bottom-heater devices, edge/top electrode structures were defined on the device in a fourth e-beam lithography step using negative-tone ARN~7520 resist. Following development, the patterns were transferred by reactive ion etching (RIE). For the top-heater devices, the devices were patterned adjacent to the transistor drain rather than directly on the drain. Following sidewall encapsulation with W, edge/top electrodes were defined on the PCM structures, and the devices were encapsulated with a \SI{50}{nm} SiO$_2$ layer deposited by atomic layer deposition (ALD). Subsequently, vias to define the heater region on the PCM stack and to the transistor drain were opened using an additional e-beam lithography step with PMMA resist. After pattern transfer by RIE and resist removal, the wafer was transferred to a physical vapor deposition (PVD) chamber, where the exposed PCM sidewalls and drain-contact surfaces were cleaned using Ar plasma. The wafer was then transferred to the ALD system, and the vias were conformally filled with \SI{60}{nm} TiN to form the top-heater structures. In a subsequent e-beam lithography step employing ARN~7520 resist, contact openings between the TiN heater and the drain electrode were defined and transferred by RIE, thereby electrically connecting the heater to the drain contact. Following the formation of both bottom-heater and top-heater structures, all devices were encapsulated with a \SI{50}{nm} SiO$_2$ layer deposited by ALD. Contact vias to the edge/top electrodes, source, gate, and ground pads were opened using e-beam lithography with PMMA resist. The developed patterns were transferred by RIE. Prior to metallization, the exposed electrodes and contact pads were cleaned \emph{in situ} using Ar plasma in the sputtering chamber. Subsequently, a \SI{150}{nm} W layer was deposited by PVD to form the contact metallization. To define the $10 \times 10$ crossbar arrays and connect the corresponding gate, source, drain, and ground terminals to dedicated routing pads, an additional e-beam lithography step was performed using ARN~7250 resist. After development, the patterns were transferred by RIE.The completed crossbar arrays were encapsulated with a \SI{70}{nm} SiO$_2$ layer deposited by RF sputtering at \SI{600}{W}. Finally, Au contact pads on W routing pads were fabricated using an optical bilayer resist process consisting of LOR~5B and AZ~1512. Following resist development, contact openings to the underlying W routing pads were created by RIE. The wafer was then transferred to the PVD chamber, where the exposed electrode surfaces were cleaned using Ar plasma. A \SI{5}{nm} Ti adhesion layer (\SI{300}{W}, \SI{20}{sccm} Ar), followed by a \SI{200}{nm} Au layer (\SI{200}{W}, \SI{20}{sccm} Ar), was deposited \emph{in situ} and subsequently patterned by lift-off to form the final contact pads. The devices were fabricated with two different $r_\mathrm{PCM}$ spans, ~\unit[200]{nm}and ~\unit[700]{nm}. \\

\subsection*{TEM studies}
\noindent Transmission electron microscopy (TEM) lamellae were prepared by focused ion beam (FIB) - scanning electron microscope (SEM), milling in a Thermo Scientific™ Helios™ 5 UC DualBeam system using 30 keV Ga+ ions, followed by low-energy (2 keV Ga+) polishing to minimize FIB-induced amorphization. A JEOL ARM200F Cs-corrected TEM, equipped with a cold-field-emission gun and operating at \unit[200]{keV}, was used to analyze the TEM lamellae. Micrographs were acquired in Z-contrast mode by High-Angle Annular Dark Field scanning TEM (HAADF–STEM). A GIF Quantum ER system was used for Electron Energy Loss Spectroscopy (EELS) measurements in Spectrum Imaging (SI) mode. Low-magnification Bright Field (BF) STEM were acquired by the STEM module of the FIB-SEM at \unit[30]{keV}.

\subsection*{Electrical characterization}
\noindent The electrical measurements were conducted using a custom-built multi-probe probe station. DC measurements of the device state were performed using a Keithley 2606B System SourceMeter. AC signals were applied to the device with a Tektronix AFG 31102 arbitrary function generator. A Tektronix oscilloscope (DPO 5104B), recorded the voltage pulses applied to and transmitted by the device. Switching between the DC and AC measurement circuits was achieved using mechanical relays in a Keithley 707A switching matrix mainframe equipped with two Keithley 7173-50 matrix cards. A custom-made relay board allowed a \unit[50]{$\Omega$} resistor to be connected from each of the 24 outputs to ground for impedance matching during programming. The chip stage was heated using a light bulb as the heating element, driven by a DC power supply (EA-PS 3150-04B) and controlled by a temperature controller (Eurotherm 2416). The programming curves were obtained by applying voltage pulses in the sequence RESET-READ-SET-READ. The RESET pulse used leading and trailing edges of \unit[8]{ns}, and a width of \unit[50]{ns}, while the amplitude was varied to create amorphous discs of increasing size. The SET pulse had fixed amplitude with leading and trailing edges of \unit[500]{ns} and \unit[5000]{ns}, and a width of \unit[10]{ns}. For MVM operations, the weights are programmed using an iterative write-and-verify scheme, targeting a conductance accuracy within a convergence tolerance of 1 \%. All READ operations were performed at \unit[0.1]{V}. The activation energies of the amorphous, crystalline, device, and liner contributions,
\(E_{\mathrm{a,amor}}\), \(E_{\mathrm{a,cry}}\), \(E_{\mathrm{a,device}}\), and
\(E_{\mathrm{a,liner}}\), were extracted from linear Arrhenius fits of the
resistance vs. temperature data according to
$\log R(T) = \log R_\infty + \frac{E_{\mathrm{a}}}{k_{\mathrm{B}}}\frac{1}{T}$,
where \(R(T)\) is the resistance at temperature \(T\), \(R_\infty\) is the
pre-exponential resistance factor, and \(k_{\mathrm{B}}\) is the Boltzmann constant.
The activation energy was obtained from the slope of the fit. The crystallization speed measurements shown in Figure~2 were performed by sweeping the pulse width and amplitude of box-shaped pulses. The trailing and falling edges for the pulse was \unit[8]{ns}. The BOX pulse width varied from \unit[10 to 1200]{ns}. To extract the drift coefficient, the resistance vs. time traces were fitted using the resistance drift relation
$R(t) = R_0 \left(\frac{t}{t_0}\right)^\nu .$
Equivalently, the drift coefficient is given by
$\nu =
\frac{\log\left(R(t)/R_0\right)}
{\log\left(t/t_0\right)}.$
Here, \(R_0\) is the device resistance at the reference time \(t_0\), and \(R(t)\) is the resistance at time \(t\).

\section*{References}
\def\url#1{}
\bibliographystyle{ieeetr}
\bibliography{References}

\newpage

\section*{Data availability}
\noindent The data that support the plots within this paper and other findings of this study are available from the corresponding author upon reasonable request.

\section*{Competing interests}
\noindent The authors declare no competing interests. 

\section*{Author contributions}
\noindent G.S.S. designed the study, and developed the models. L.C. performed experiments on the crossbar arrays, and contributed to data fitting. V.J.O fabricated the devices and integrated on crossbar arrays. A.M.M performed the TEM studies. A.R. and M.B. contributed with wafers. A.S. provided technical and management support. 

\section*{Acknowledgments}

\noindent This work was supported by the European Research Council Grant INFUSED (Grant No. 101222715) and by the IBM Research AI Hardware Center. We thank Timothy Philicelli, Stephan Menzel, Siddharth Gautam, Andrea Cassini, Jesse Luchtenveld for valuable technical discussions and contributions, and Vijay Narayanan for his management support. We also acknowledge support from Urs Egger toward electrical characterization setup. We also acknowledge the cleanroom operations team at the Binnig and Rohrer Nanotechnology Center for their technical support, as well as the administration of CNR-IMM for their assistance.

\clearpage

\section*{Figures}
%------------------------------------------------------
% Figure 1
%------------------------------------------------------

%------------------------------------------------------
% Figure 2
%------------------------------------------------------

%------------------------------------------------------
% Figure 3
%------------------------------------------------------

%------------------------------------------------------
% Figure 4
%------------------------------------------------------

\clearpage

\end{document}